%% file: main.tex
\documentclass[conference]{IEEEtran}
\IEEEoverridecommandlockouts

\usepackage{amssymb}
\usepackage{svg}
\usepackage{amsthm}
\usepackage{graphicx}
\usepackage{subcaption}
\usepackage[font=footnotesize]{caption}
\usepackage[cmex10]{amsmath}
\usepackage{booktabs}
\usepackage{enumitem}
\usepackage{url}
\usepackage[hang,flushmargin]{footmisc}
\usepackage{cite}
\usepackage{geometry}
\usepackage{dblfloatfix}
\usepackage{xcolor}

\newcommand{\ID}[2]{I_\mathrm{D}^{\mathrm{#1}{#2}}}
\newcommand{\VD}[2]{V_\mathrm{D}^{\mathrm{#1}{#2}}}
\newcommand{\ISH}[2]{I_\mathrm{SH}^{\mathrm{#1}{#2}}}
\newcommand{\IPH}[2]{I_\mathrm{PH}^{\mathrm{#1}{#2}}}
\newcommand{\VPV}[2]{V_\mathrm{PV}^{\mathrm{#1}{#2}}}
\newcommand{\IS}[2]{I_\mathrm{S}^{\mathrm{#1}{#2}}}
\newcommand{\IOUT}[2]{I_\mathrm{out}^{\mathrm{#1}{#2}}}
\newcommand{\Ibypass}[2]{I_\mathrm{by}^{\mathrm{#1}{#2}}}
\newcommand{\RS}[2]{R_\mathrm{S}^{\mathrm{#1}{#2}}}
\newcommand{\RSH}[2]{R_\mathrm{SH}^{\mathrm{#1}{#2}}}
\newcommand{\ZLoad}[1]{Z_\mathrm{load}^\mathrm{#1}}
\newcommand{\Iarray}[1]{\IOUT{a}{}}
\newcommand{\Varray}[1]{\VPV{a}{}}
\newcommand{\SDM}[1]{\mathrm{SDM}_{\mathrm{#1}}}
\newcommand{\PPDM}[1]{\mathrm{PPDM}_{\mathrm{A}}}
\newcommand{\node}[1]{\textit{n}_\mathrm{#1}^{\mathrm{p}(i,j)}}

\newif\ifshowchanges
\showchangesfalse
\ifshowchanges
    \newcommand{\rev}[1]{\textcolor{blue}{#1}}
    \newenvironment{changed}{\begingroup\color{blue}}{\endgroup}
\else
    \newcommand{\rev}[1]{#1}
    
\fi

\title{\LARGE \bf Circuit-Based Analysis of Per-Panel Diode Model for Photovoltaic Array}
\author{Peng Sang$^\dagger$, Santhosh Balasubramanian$^\ddagger$, Amritanshu Pandey$^\dagger$
\vspace{-0.5cm}
\thanks{$^\dagger$Peng and Amritanshu are with the Department of Electrical and Biomedical Engineering, University of Vermont, Burlington, VT, U.S. Email: {\tt peng.sang;amritanshu.pandey@uvm.edu}}
\thanks{$^\ddagger$Santhosh is with the Davidson School of Chemical Engineering, Purdue University, Lafayette, IN, U.S. Email: {\tt balasu41@purdue.edu}}}

\begin{document}
\date{}
\maketitle
\vspace{-8mm}
\pagenumbering{gobble}

\begin{abstract}
\input{Abstract}
\end{abstract}
\vspace{2mm}

\section*{Nomenclature}
\input{Nomenclature}

\section{Introduction}
\input{Introduction}

\section{Aggregated single diode model for array}\label{sec:SDMP_SDMA}
\input{Preliminaries}

\section{Per-panel diode model for array}\label{sec:PPDMA}
\input{Problem_description}

\section{Experimental Setup}\label{sec:experiment}
\input{Experiment_setup}

\section{Numerical Results}\label{sec:Numerical Results}
\input{Experiment}

\section{Conclusions}
\input{Conclusion}

\newpage
\bibliographystyle{IEEEtran}
\bibliography{Ref}
\end{document}

%% file: Abstract.tex
Photovoltaic (PV) systems now provide grid services such as peak shaving and demand response.
To support these services, the current practice is to represent a PV array with an aggregated single-diode model ($\SDM{A}$).
Due to its abstraction, $\SDM{A}$ cannot capture PV behaviors under non-uniform real-world conditions such as partial shading and hotspots.
Therefore, we develop a per-panel circuit-based modeling framework for PV arrays that achieves a desirable trade-off between model fidelity and available sensing.
Also, we establish conditions under which the proposed per-panel array circuit model is mathematically equivalent to the aggregated single-diode array model.
Simulation results with real-world model parameters show that the per-panel diode model more accurately represents array electrical behavior under non-ideal conditions.
In maximum power point tracking control, the per-panel diode model improves power estimation accuracy by 17.2\% under partial shading and 5.5\% under hotspot conditions.

%% file: Nomenclature.tex
\addcontentsline{toc}{section}{Nomenclature}

\begin{description}[align=left, labelwidth=1cm, labelsep=1em]
\hrule
\vspace{2mm}
  \item[\(m^{\mathrm{c}}\)] Number of cells in series in a panel
  \item[\(m^{\mathrm{p}}\)] Number of panels in series in an array
  \item[\(n^{\mathrm{c}}\)] Number of cells in parallel in a panel
  \item[\(n^{\mathrm{p}}\)] Number of panels in parallel in an array
 \item[\(i, j\)] Row and column index of panel \((i,j)\) in an array
  \item[\(q\)] Elementary electric charge (C)
  \item[\(\eta\)] Diode ideality factor
  \item[\(k\)] Boltzmann constant (J/K)
  \item[\(T\)] Cell temperature (K)
  \item[$\IPH{}{}$] Photocurrent (A)
  \item[$\ID{}{}$] Diode current (A)
  \item[\(I_0\)] Reverse saturation current of the diode (A)
  \item[$\RS{}{}$] Series resistance (\(\Omega\))
  \item[$\IS{}{}$] Current through series resistance (A)
  \item[$\RSH{}{}$] Shunt resistance (\(\Omega\))
  \item[$\ISH{}{}$] Current through shunt resistance (A)
  \item[$\IOUT{}{}$] Output terminal current (A)
  \item[$\ZLoad{}{}$] Inverter load impedance (\(\Omega\))
  \item[$\VD{}{}$] Voltage across photodiode (V)
  \item[$\VPV{}{}$] Voltage across SDM terminal (V)
  \item[$\VPV{a}{}$] Voltage across array (V)
  \item[$\IOUT{a}{}$] Current output of the array (A)
  \item[$\mathrm{SDM}_\mathrm{P,A}$] Single diode model of panel/array
  \item[$\mathrm{PPDM}_\mathrm{A}$] Per-panel diode model of array
\vspace{1mm}
\hrule\hrule
\end{description}

%% file: Introduction.tex
\noindent\textbf{{Motivation}}:
\noindent
As solar PV penetration grows~\cite{IRENA_2024,IRENA2019a}, solar PV will provide essential services beyond carbon-free energy, including peak shaving and demand response.
In the New England region, roughly one-third of days are characterized as \textit{duck curve days} \cite{MollyEnking_2025}, which makes accurate modeling of solar variability essential.
As such, we need \textit{practical} yet \textit{accurate} models for solar PV for analysis of various services.
The state-of-the-art models do not accurately characterize the behavior of a non-uniform PV array for practical analysis, especially under system variability due to shading and hotspot conditions.
For instance, consider a solar PV facility in Vermont with a weather station and per-panel measurements.
For this facility, we observe a severe power mismatch between the simulation output of the aggregated single-diode array model and the measurements.
This discrepancy reveals that the single-diode array model lacks the fidelity to represent non-uniform arrays.
It motivates us to identify the conditions under which the aggregated model remains a lossless aggregation, and the scenarios where higher-fidelity representations are required to capture non-uniform behavior.

\noindent \textbf{{{State-of-the-art and limitations}}}:\label{sec:SOA_limitation}
The commonly used PV array model is the five-parameter single diode model ($\SDM{A}$), which generalizes panel-level behavior to the array level~\cite{hua1998study,gutierrez2024modeling}.
The aggregation rule of the generalized PV array model is shown in Table~\ref{tab:agg_rules}; this rule assumes identical panel parameters and uniform conditions across the array.
Under this assumption, the $\SDM{A}$ accurately represents the behavior of utility-scale PV systems~\cite{tsai2008development,fang2020active}.
However, the assumptions of uniformity do not always hold in the field; non-uniform effects cause inconsistency between $\SDM{A}$ and reality~\cite{Nguyen2015}.
Recent work, such as~\cite{ccelik2025reconfigured}, improves the modeling of PV cells/panels by incorporating a small resistance in series with the diodes in single- and double-diode models.
But the improved models still inherit the same uniformity assumptions.
Existing studies (e.g.,~\cite{pendem2018modelling}) simulate per-panel behavior and demonstrate the impact of non-uniform conditions through simulations or experiments, but lack a formal mathematical modeling framework to study these models.

\noindent\textbf{{{Insights of the problem}}}:
As the key insight, we find that a per-panel model for a PV array offers a good trade-off between higher sensing and model fidelity.
Per-cell models offer even higher fidelity, but acquiring cell-level measurements often necessitates additional equipment.
The panel level is generally the smallest unit that plant operators observe and validate during operation.
PV panel parameters are readily accessible from manufacturer spec sheets, while cell-level parameters are generally not.
We also find that there is a lack of a mathematical framework that refers to the underlying theory for constructing panel-level PV array models.
Thus, no existing work derives the conditions under which the aggregated single diode model ($\SDM{A}$) is equivalent to the full per-panel diode model ($\PPDM{}$).
The key contributions of this work are:
\begin{itemize}
    \item Development of an equivalent circuit model and analysis of the  $\PPDM{A}$ of a PV array
    \item Derivation of necessary conditions for mathematical equivalence between the $\PPDM{A}$ and the $\SDM{A}$ of a PV array
    \item Comparison of performance between $\PPDM{A}$ and $\SDM{A}$, with real PV array system parameters
\end{itemize}

%% file: Preliminaries.tex
\noindent It is common practice to represent both PV panels and arrays using the five-parameter single diode model (SDM).
To illustrate that the two aggregations are built on the same uniform conditions, we discuss both aggregated $\SDM{P}$ from the individual cell (\ref{sbsec:PVSDM}) and aggregated $\SDM{A}$ from the individual panel (\ref{sbsec:SDMA}).
\subsection{Single diode model of PV panel}\label{sbsec:PVSDM}
\noindent Previous work~\cite{osti_1048979} derives the equations to build $\SDM{P}$ from individual PV cells.
It assumes that all cells in a panel are identical, as they are in close proximity and part of the same panel package, therefore share the same specifications.
It also assumes that all cells in a panel share the same irradiance and temperature.

In most PV models, the diode behavior is captured by the Shockley diode equation:
\begin{equation} \label{eq:Shockley Equation}
\ID{c}{} = I_0^\mathrm{c} \left[ \exp\left(\frac{q\VD{c}{}}{\eta kT}\right) - 1 \right]
\end{equation}
Under the above assumptions, for a panel that has $m^\mathrm{c}$ and $n^\mathrm{c}$ cells in parallel and series , \cite{Nguyen2015} represents the panel diode voltage $\VD{p}{}$ as a function of cell diode voltage $\VD{c}{}$:
\begin{equation}\label{eq:c_p_VD}
    \VD{p}{} = m^\mathrm{c}\VD{c}{}
\end{equation}
\noindent and diode current $\ID{p}{}$ in panel $\SDM{P}$ as a function of diode current $\ID{c}{}$ in cell $\SDM{P}$:
\begin{equation}\label{eq:c_p_ID}
    \ID{p}{} = n^\mathrm{c}\ID{c}{}
\end{equation}
\noindent as the cells are connected in series and parallel.
Further, \cite{jereminov2016improving} abbreviates the Shockley Equation for the panel as:
\begin{equation} \label{eq:idp}
\ID{p}{} = I_0^\mathrm{p} \left[ \exp\left(\frac{\VD{p}{}}{\alpha^\mathrm{p}}\right) - 1 \right]
\end{equation}
\noindent where:
\begin{equation}\label{eq:c_p_I0}
    I^\mathrm{p}_0 = n^\mathrm{c}I^\mathrm{c}_0
\end{equation}
\noindent and for a single cell:
\begin{equation} \label{eq:abb_alpha}
\alpha^\mathrm{c} = \frac{\eta kT}{q}
\end{equation}
\noindent and for a panel that has $m^\mathrm{c}$ cells connected in series:
\begin{equation}\label{eq:alpha_temp}
    \alpha^\mathrm{p} = m^\mathrm{c}\alpha^\mathrm{c} = \frac{\eta  kTm^\mathrm{c}}{q}
\end{equation}
\noindent where
$I_0^\mathrm{c}$ and $I_0^\mathrm{p}$ are the reverse saturation currents of the cell and panel SDM.
$\alpha^\mathrm{c}$ and $\alpha^\mathrm{p}$ are the thermal voltages of the cell and panel SDM, $q$ is the absolute value of the electric charge of an electron, $T$ is the cell temperature in Kelvin, and $k$ is the Boltzmann constant; $m^\mathrm{c}$ is the number of cells connected in series.
For the photon current relationship between the panel $\SDM{}$ model and cell $\SDM{C}$ model, we have:
\begin{equation}\label{eq:c_p_IPH}
    \IPH{p}{} =n^\mathrm{c}\IPH{c}{}
\end{equation}

In $\SDM{P}$, the series resistance parameter $\RS{p}{}$ models the contact resistance between silicon and electrode surfaces of the cells, the resistance of electrodes, and the current flow resistance in the cable connections.
Its relationship as a function of cell $\SDM{C}$ parameter is given by:
\begin{equation} \label{eq:c_p_RS}
    \RS{p}{} = \frac{m^\mathrm{c}}{n^\mathrm{c}}\RS{c}{}
\end{equation}
\noindent Subsequently, the current through series resistance $\IS{p}{}$ in panel $\SDM{P}$ is:
\begin{equation}\label{eq:c_p_IS}
    \IS{p}{} = n^\mathrm{c}\IS{c}{}
\end{equation}
\noindent and the terminal voltage of a panel should be:
\begin{equation}\label{eq:c_p_VPV}
    \VPV{p}{} = m^\mathrm{c}\VPV{c}{}
\end{equation}

To model the leakage current of the P–N junction, an aggregated shunt resistance \(\RSH{p}{}\) represents the combined leakage current of all cells, where:
\begin{equation} \label{eq:c_p_RSH}
    \RSH{p}{} = \frac{m^\mathrm{c}}{n^\mathrm{c}}\RSH{c}{}
\end{equation}

\begin{figure}
    \centering
    \includegraphics[width=0.8\linewidth]{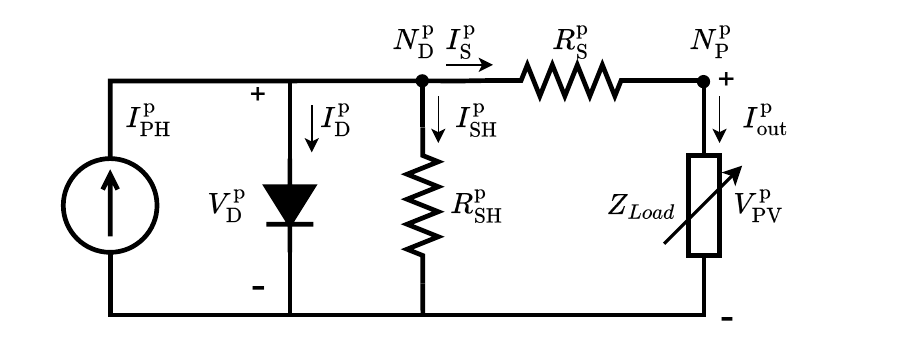}
    \caption{Schematic of $\SDM{P}$.}
    \label{fig:Schematic of Single Solar Cell}
\end{figure}

The resultant model is the $\SDM{P}$ or five-parameter model for representing the behavior of a PV panel \cite{Jordehi2016}.
We show the schematic of the $\SDM{P}$ in Fig.~\ref{fig:Schematic of Single Solar Cell}.
The $\SDM{P}$ model describes the panel physics with voltage-current relationships, given five aggregated parameters. We derive these V-I relationships for the $\SDM{P}$ based on circuit fundamentals: Kirchhoff's laws and Ohm's law. Writing KCL equations for node $N_\mathrm{D}^\mathrm{p}$ and $N_\mathrm{P}^\mathrm{p}$:
\begin{subequations}\label{eq:Solar Panel Basic eqns}
\begin{equation}\label{eq:Node 1}
    \begin{split}
    \text{node}\;N_\mathrm{D}^\mathrm{p}: \IPH{p}{} - \ID{p}{} - \ISH{p}{} - \IS{p}{} = 0
    \end{split}
\end{equation}
\begin{equation}\label{eq:Node 2}
    \begin{split}
    \text{node}\;N_\mathrm{P}^\mathrm{p}:\IS{p}{} - \IOUT{p}{} = 0
    \end{split}
\end{equation}
\end{subequations}

\noindent where the diode current \(\ID{p}{}\) is given by \eqref{eq:idp}:

The expressions for \(\ISH{p}{}\) and \(\IS{p}{}\) in terms of \(\VD{p}{}\) and \(\VPV{p}{}\) can be derived using Ohm's Law:
\begin{subequations}\label{eq:Ohms law across RSH and RS}
\begin{equation}\label{eq:RSH Ohm Law}
    \begin{split}
    \frac{\VD{p}{}}{\RSH{p}{}} - \ISH{p}{} = 0
    \end{split}
\end{equation}
\begin{equation}\label{eq:RS Ohm Law}
    \begin{split}
    \frac{\VD{p}{}-\VPV{p}{}}{\RS{p}{}} - \IS{p}{} = 0
    \end{split}
\end{equation}
\end{subequations}
\noindent Ohm's Law across grid equivalent impedance \(\ZLoad{}\):
\begin{equation} \label{eq:ZLoad}
   \frac{\VPV{p}{}}{ \ZLoad{}} - \IOUT{p}{} = 0
\end{equation}
\noindent Substituting equations~\eqref{eq:idp}, \eqref{eq:Node 2}, and~\eqref{eq:RSH Ohm Law} into \eqref{eq:Node 1} we get \eqref{eq:Final Equation 1 for Solar cell}.
Substituting equations~\eqref{eq:RS Ohm Law} into \eqref{eq:Node 2}, we get \eqref{eq:Final Equation 2 for Solar cell}.
\begin{subequations}\label{eq:Solar Panel Characteristics}
\begin{equation}\label{eq:Final Equation 1 for Solar cell}
    \begin{split}
    \IPH{p}{} - I_0^\mathrm{p} \left[ \exp\left(\frac{\VD{p}{}}{\alpha^\mathrm{p}}\right) - 1 \right] - \frac{\VD{p}{}}{\RSH{p}{}} - \IOUT{p}{} = 0
    \end{split}
\end{equation}
\begin{equation}\label{eq:Final Equation 2 for Solar cell}
    \begin{split}
    \frac{\VD{p}{}-\VPV{p}{}}{\RS{p}{}} - \IOUT{p}{} = 0
    \end{split}
\end{equation}
\end{subequations}
\noindent Equations \eqref{eq:Final Equation 1 for Solar cell} and \eqref{eq:Final Equation 2 for Solar cell}, along with the control equation that determines the grid impedance ($\ZLoad{p}$) seen by the DC source, map the complete physics of the PV panel.

\subsection{Single diode model of PV array}\label{sbsec:SDMA}
\noindent
The same aggregation process used in \ref{sbsec:PVSDM} to build $\SDM{P}$ extends similarly from panels to arrays.
For a PV array with $m^\mathrm{p}$ rows and $n^\mathrm{p}$ columns of panels, $\SDM{A}$ represents its lumped behavior under uniform conditions.
We omit the step-by-step derivation as it mirrors \ref{sbsec:PVSDM}.
Applying the analogous KCL and Ohm's law derivation to the panel parameters yields a similar set of equations for $\SDM{A}$:
\begin{equation}\label{eq:eq1_for_AAM}
    \begin{split}
    \IPH{a}{} - I_0^\mathrm{a} \left[ \exp\left(\frac{\VD{a}{}}{\alpha^\mathrm{a}}\right) - 1 \right] - \frac{\VD{a}{}}{\RSH{a}{}} - \IOUT{a}{} = 0
    \end{split}
\end{equation}
\begin{equation}\label{eq:eq2_for_AAM}
    \begin{split}
    \frac{\VD{a}{}-\VPV{a}{}}{\RS{a}{}} - \IOUT{a}{} = 0
    \end{split}
\end{equation}
\begin{equation}\label{eq:eq3_for_AAM}
    \begin{split}
    \frac{\VPV{a}{}}{\ZLoad{}} - \IOUT{a}{} = 0
    \end{split}
\end{equation}

\begin{table}[t]
\caption{Aggregation Rules for an $m^{\mathrm{p}} \times n^{\mathrm{p}}$ PV Array}
\label{tab:agg_rules}
\centering
\renewcommand{\arraystretch}{1.3}
\setlength{\tabcolsep}{5pt}
\begin{tabular}{@{}lcl@{}}
\toprule
\textbf{Parameter group} & \textbf{Scaling} & \textbf{Expression} \\
\midrule
Thermal voltage          & $m^{\mathrm{p}}$              & $\alpha^{\mathrm{a}} = m^{\mathrm{p}} \alpha^{\mathrm{p}} = \eta k T m^{\mathrm{p}} / q$ \\
Voltages $(V_D,\,V_{\mathrm{PV}})$ & $m^{\mathrm{p}}$              & $V^{\mathrm{a}} = m^{\mathrm{p}} V^{\mathrm{p}}$ \\
Currents $(I_0,\,I_{\mathrm{PH}},\,I_{\mathrm{SC}})$ & $n^{\mathrm{p}}$ & $I^{\mathrm{a}} = n^{\mathrm{p}} I^{\mathrm{p}}$ \\
Resistances $(R_S,\,R_{\mathrm{SH}})$ & $m^{\mathrm{p}}/n^{\mathrm{p}}$ & $R^{\mathrm{a}} = (m^{\mathrm{p}}/n^{\mathrm{p}}) R^{\mathrm{p}}$ \\
\bottomrule
\end{tabular}
\vspace{2pt}
\begin{flushleft}
\footnotesize
$I^{\mathrm{a}}_0$, $I^{\mathrm{p}}_0$: reverse saturation currents (array, panel).
$\alpha^{\mathrm{a}}$, $\alpha^{\mathrm{p}}$: thermal voltage.
$q$: electron charge; $T$: cell temperature (K); $k$: Boltzmann constant.
\end{flushleft}
\end{table}

The aggregation rules for generalizing panel-level behavior to the array-level single-diode model ($\SDM{A}$) are summarized in Table~\ref{tab:agg_rules}.

%% file: Problem_description.tex
\begin{figure}
    \centering
    \includegraphics[width=0.85\linewidth]{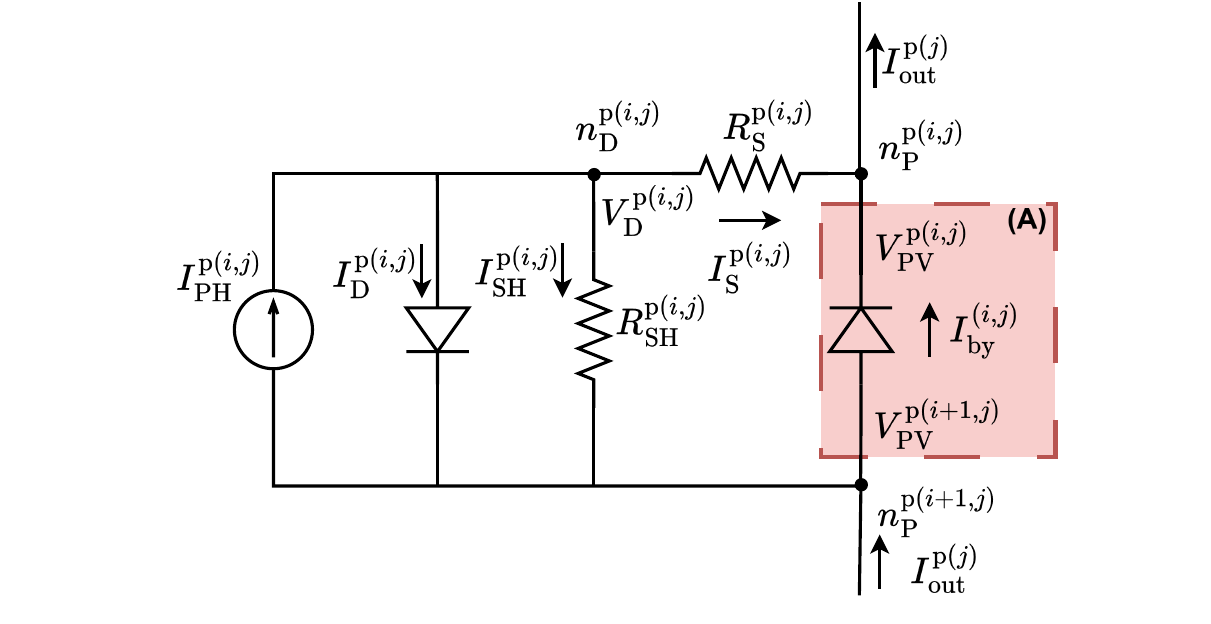}
    \caption{$\SDM{p}$ circuit of a panel $(i, j)$ in array; bypass diode shown in (A) is optional.
    Each panel has two nodal voltages $\VD{p}{(i,j)}$ and $\VPV{p}{(i,j)}$ corresponding to nodes $\node{D}$ and $\node{P}$, respectively.
    }
    \label{fig:PV_farm}
\end{figure}

\noindent This section introduces a per-panel diode model ($\PPDM{A}$) and derives the conditions for mathematical equivalence between $\PPDM{A}$ and $\SDM{A}$.
\begin{figure}
    \centering
    \includegraphics[width=0.85\linewidth]{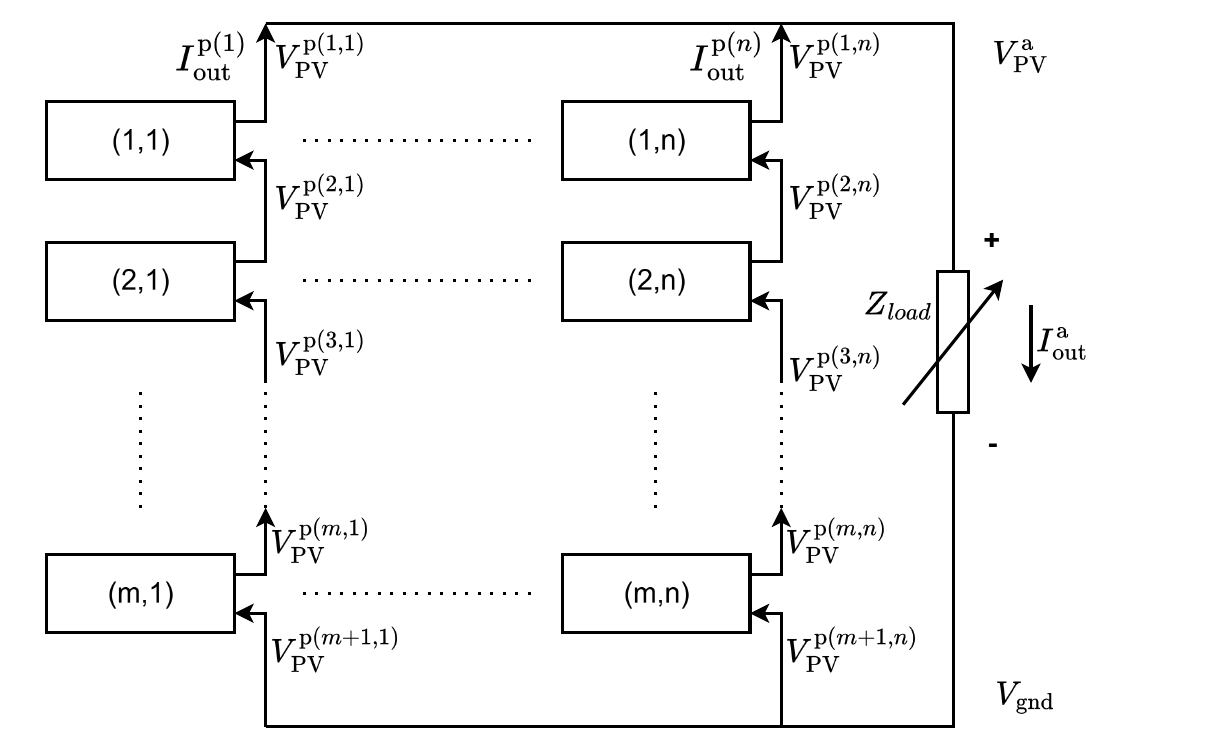}
    \caption{Simplified PV-array block diagram showing array-level variables $(\Varray{},\Iarray{})$, panel nodal voltage $\VPV{p}{}$, string output current $\IOUT{p}{}$, and equivalent grid impedance $\ZLoad{}$, each block represents a panel modeled via $\SDM{P}$.}
    \label{fig:PV_Farm_Simple}
\end{figure}
We show the schematic of a PV array in Fig. \ref{fig:PV_Farm_Simple}.
We assume uniform cell parameters within each panel based on assumptions of available sensing.
The current output of the array is the sum of the current output of each string:
\begin{equation}\label{eq:Iarr_sumIout}
    \Iarray{}  = \sum^n_{j = 1}\IOUT{p}{(j)} = \frac{\Varray{} - V_\text{gnd}}{\ZLoad{}}
\end{equation}
We ground the negative terminal of the last panel in all strings ($j = 1,\dots,n$) \rev{and represent the ground potential as $V_\text{gnd}$. 
We then} define the array terminal voltage $\Varray{}{}$ as a function of voltages from the first panel of a string:
\begin{subequations}
    \begin{align}
        \VPV{p}{(1,j)} - \Varray{} = 0\;\;\forall j \in \{1,2,\dots,n\}\label{eq:Varr}\\
        \VPV{p}{(m+1,j)} - V_\text{gnd} = 0 \;\;\forall j \in \{1,2,\dots,n\}\label{eq:Vgnd}
    \end{align}
\end{subequations}\
Next, we zoom in on one panel indexed by $(i,j)$ within the array and examine its $\SDM{P}$ model (see schematic in Fig. \ref{fig:PV_farm}). The KCL equations for panel $(i, j)$ are given by:
\begin{subequations}\label{eq:PPDM_node_eqs}
    \begin{align}
        &\text{node}\;\node{D}:\IPH{p}{(i,j)} - \ID{p}{(i,j)} - \ISH{p}{(i,j)} - \IS{p}{(i,j)} = 0 \label{eq:PPDM_node_1} \\
        &\text{node}\;\node{P}:\IS{p}{(i,j)}- \IOUT{p}{(j)} = 0 \label{eq:PPDM_node_2}
    \end{align}
\end{subequations}

\noindent Following from \eqref{eq:PPDM_node_1}, and \eqref{eq:PPDM_node_2} the set of KCL is equivalent to:
\begin{subequations}\label{eq:perpanel_final_eqs}
    \begin{align}
   &\text{node}\;\node{D}:\notag \\ &\IPH{p}{(i,j)}
    - I_0^{\mathrm{p}{(i,j)}} \left( \exp\big[\frac{\VD{p}{(i,j)} - \VPV{p}{(i+1,j)}}{\alpha^{\mathrm{p}{(i,j)}}}\big] - 1 \right) \notag \\
    & - \frac{\VD{p}{(i,j)} - \VPV{p}{(i+1,j)}}{\RSH{p}{(i,j)}} - \frac{\VD{p}{(i,j)} - \VPV{p}{(i,j)}}{\RS{p}{(i,j)}}  = 0 \label{eq:perpanel_final_eq_1}\\
     &\text{node}\;\node{P}:\frac{\VD{p}{(i,j)} - \VPV{p}{(i,j)}}{\RS{p}{(i,j)}} - \IOUT{p}{(j)} = 0  \label{eq:perpanel_final_eq_2}\\
     &\quad \forall i \in \{1, \dots, m\}; \forall j \in \{1, \dots, n\}\notag
\end{align}
\end{subequations}

\noindent To include a bypass diode in the model, we add the following circuit equation:
\begin{subequations}
\begin{equation}
        \IS{p}{(i,j)} - \IOUT{p}{(j)} + \Ibypass{}{(i,j)} = 0 \label{eq:farm_node_2_w_diode}
\end{equation}
\begin{equation}\label{eq:bypass_diode_current}
\begin{split}
    \Ibypass{}{(i,j)} - I_0^{\text{by}{(i,j)}}(\exp[\frac{\VPV{p}{(i+1,j)} - \VPV{p}{(i,j)}}{\alpha^{\text{by}{(i,j)}}}]-1)  = 0\\ \quad \forall i \in \{1, \dots, m\}; \forall j \in \{1, \dots, n\}
\end{split}
\end{equation}
\end{subequations}
\noindent \eqref{eq:node_2_w_diode} will replace \eqref{eq:perpanel_final_eq_2} when the bypass diode is included.
Equation \eqref{eq:node_2_w_diode} is obtained by substituting \eqref{eq:bypass_diode_current} into \eqref{eq:farm_node_2_w_diode} and replaces \eqref{eq:perpanel_final_eq_2} in the bypass-diode case:
\begin{equation}\label{eq:node_2_w_diode}
\begin{split}
    &\frac{\VD{p}{(i,j)} - \VPV{p}{(i,j)}}{\RS{p}{(i,j)}} - \IOUT{p}{(j)} \\
    &+ I_0^{\text{by}{(i,j)}}(\exp[\frac{\VPV{p}{(i+1,j)} - \VPV{p}{(i,j)}}{\alpha^{\text{by}{(i,j)}}}]-1) = 0 \\
    & \forall i \in \{1, \dots, m\}; \forall j \in \{1, \dots, n\}
\end{split}
\end{equation}
With both array and panel equations, we can solve for each node voltage in the array with the equation set \eqref{eq:Iarr_sumIout},\eqref{eq:Vgnd},\eqref{eq:Varr},\eqref{eq:perpanel_final_eq_1},\eqref{eq:perpanel_final_eq_2} (without bypass diode) or \eqref{eq:Iarr_sumIout},\eqref{eq:Vgnd},\eqref{eq:Varr},\eqref{eq:perpanel_final_eq_1},\eqref{eq:node_2_w_diode} (with bypass diode), given $\SDM{P}$ parameters for each panel and $\ZLoad{}$.
$\PPDM{A}$ follows equations in \cite{osti_1048979} to account for the change of parameters under different temperature and irradiance conditions.
$\ZLoad{}$ mimics a controlled current source and is a function of the inverter control strategy.

Next, we prove by \textit{contradiction} that the aggregation of the panel model is lossless (i.e. $\PPDM{A}$ is mathematically equivalent to $\SDM{A}$) only when all panels in the array are uniform (they share the same five parameters).
\begin{proof}
We examine the conditions under which the terminal \textit{I-V} characteristics, that is, array output current and terminal voltage, of the $\SDM{A}$ and $\PPDM{A}$ are mathematically equivalent.

\noindent \textbf{Claim.} Array current and terminal voltage in $\SDM{A}$, see \eqref{eq:eq1_for_AAM}-\eqref{eq:eq3_for_AAM}, are mathematically equivalent to array current and terminal voltage in  $\PPDM{A}$, respectively, when the various panels in an array have a non-uniform set of 5 parameters.

From the Claim, as part of the \textit{I-V} characteristic, the two models' output currents are equal and satisfy the following relationships:
\begin{equation}\label{eq:claimeq}
    \IOUT{a}{} = \sum_{j = 1}^n\IOUT{p}{(j)}
\end{equation}
\noindent For notational convenience, we define:
\begin{subequations}
\begin{align}
    x^{\mathrm{a}} &= \frac{V_D^{\mathrm{a}}}{\alpha^{\mathrm{a}}} \\
    x^{\mathrm{p}{(i,j)}} &= \frac{\VD{p}{(i,j)} - \VPV{p}{(i+1,j)}}{\alpha^{\mathrm{p}{(i,j)}}}
\end{align}
\end{subequations}
\rev{where $x^{\mathrm{a}}$ and $x^{\mathrm{p}{(i,j)}}$ denote the dimensionless Shockley exponential arguments for the array and panel $(i,j)$, respectively.}
Substituting \eqref{eq:eq1_for_AAM} and \eqref{eq:PPDM_node_1}--\eqref{eq:PPDM_node_2} into the two sides of \eqref{eq:claimeq} yields
\begin{equation}\label{eq:expand}
\begin{split}
\IPH{a}{}
- I_0^{\mathrm{a}}\!\left[\exp\!\left(x^{\mathrm{a}}\right)-1\right]
- \ISH{a}{}
= {} &\\
\sum_{j=1}^{n}\IPH{p}{(i,j)}
-\sum_{j=1}^{n}\ID{p}{(i,j)}
-\sum_{j=1}^{n}\ISH{p}{(i,j)},
&\forall i \in \{1,\dots,m\}
\end{split}
\end{equation}
For the claim to hold \rev{over the entire I--V operating range}, the additive decomposition in \eqref{eq:expand} must be \emph{aligned parameter-wise.}\footnote{Throughout the proof, we exploit the fact that \eqref{eq:expand} must hold across the entire I–V operating range; linear independence of the constant, exponential, and linear in $\VD{}{}$ terms then forces each functional component to balance individually.}
Each parameter in $\SDM{A}$ must be accounted for by its \emph{counterpart parameters} in $\PPDM{A}$.
Thus, the following subequations must hold $\forall i$:
\begin{subequations}\label{eq:elemetwise}
\begin{align}
\IPH{a}{} &= \sum_{j=1}^{n}\IPH{p}{(i,j)} \label{eq:Iph_sumIph}\\
I_0^{\mathrm{a}}\!\left[\exp\!\left(x^{\mathrm{a}}\right)-1\right]
&= \sum_{j=1}^{n}\ID{p}{(i,j)}\label{eq:diodeComponent} \\
\ISH{a}{} &= \sum_{j=1}^{n}\ISH{p}{(i,j)}\label{eq:Ish_sumIsh}
\end{align}
\end{subequations}
We focus on \eqref{eq:diodeComponent} to examine the additive condition by expanding the sum on the RHS:
\begin{equation}
I_0^{\mathrm{a}}\!\left[\exp\!\left(x^{\mathrm{a}}\right)-1\right]
= \sum_{j=1}^{n} I_0^{\mathrm{p}(i,j)}\!\left[\exp\!\left(x^{\mathrm{p}(i,j)}\right)-1\right]
\end{equation}

Further, parameter alignment ensures that the saturation current ($I_0$) of the $\SDM{A}$ is equal to the sum of diode saturation currents across parallel branches of $\PPDM{}$:
\begin{subequations}
    \begin{align}
        I_0^{\mathrm{a}} \exp\left( x^{\mathrm{a}} \right) &= \sum_{j=1}^{n} I_0^{\mathrm{p}{(i,j)}} \exp\left( x^{\mathrm{p}{(i,j)}} \right) \label{eq:exp_Iout}\\
    I_0^{\mathrm{a}} &= \sum_{j=1}^{n} I_0^{\mathrm{p}(i,j)} \label{eq:I0_sumI0}
\end{align}
\end{subequations}

\noindent Dividing both sides of \eqref{eq:exp_Iout} by $\exp(x^{\mathrm{a}})$ results in:
\begin{equation} \label{eq:derive1}
    I_0^{\mathrm{a}} = \sum_{j=1}^{n} I_0^{\mathrm{p}{(i,j)}} \exp(x^{\mathrm{p}{(i,j)}} - x^{\mathrm{a}})
\end{equation}

\noindent Now substitute \eqref{eq:I0_sumI0} into \eqref{eq:derive1}:
\begin{equation}\label{eq:ID_SUM}
    \sum_{j=1}^{n} I_0^{\mathrm{p}(i,j)} = \sum_{j=1}^{n} I_0^{\mathrm{p}(i,j)} \exp(x^{\mathrm{p}(i,j)} - x^{\mathrm{a}})
\end{equation}
\noindent \rev{For the claim to hold, the additive decomposition in \eqref{eq:expand} must be \emph{aligned parameter-wise, without cross-panel compensation}}, this implies\footnote{We apply the same operating-range argument to \eqref{eq:ID_SUM}. 
The per-panel quantities ($x^{\mathrm{p}(i,j)} - x^{\mathrm{a}})$ vary independently as $\ZLoad{}$ sweeps.
This forces each exponential term to equal 1, yielding \eqref{eq:equal_exp}.}:
\begin{equation} \label{eq:equal_exp}
    \exp(x^{\mathrm{p}(i,j)} - x^{\mathrm{a}}) = 1 \quad \forall j \in \{1, \dots, n\}
\end{equation}
\begin{equation} \label{eq:x_equal}
    x^{\mathrm{p}(i,j)} = x^{\mathrm{a}} \quad \forall i \in \{1, \dots, m\}; \forall j \in \{1, \dots, n\}
\end{equation}
Returning to the original variable definitions, this implies:
\begin{equation} \label{eq:x_expanded}
\begin{split}
    \frac{\VD{p}{(i,j)} - \VPV{p}{(i+1,j)}}{\alpha^{\mathrm{p}{(i,j)}}} = \frac{\VD{a}{}}{\alpha^{\mathrm{a}}},
    \forall i \in \{1, \dots, m\}; \forall j \in \{1, \dots, n\}
\end{split}
\end{equation}
Thus, for the aggregation to be exact, the per-panel voltage and diode parameters must satisfy strict exponential alignment not only for the terminal voltage, but also for every panel index $(i, j)$. This contradicts the assumption \textit{that the panels in the array can have a non-uniform set of five parameters} stated in the Claim above. Therefore, the Claim cannot be true, and a uniformity condition is required to have a lossless aggregation.
\end{proof}
\noindent\textbf{{{Control Implementation}}}:
We evaluate $\SDM{A}$ and $\PPDM{A}$ under a maximum power point tracking (MPPT) control scheme.
For $\SDM{A}$, we find the voltage and current of the maximum power point (MPP) implicitly through inclusion of a control equation to find the $\ZLoad{}$ that will make the array operate at its maximum power point (MPP)~\cite{jereminov2016improving}:
\begin{equation}\label{eq:mpp}
    \frac{\partial(\IOUT{a}{}\VPV{a}{})}{\partial\VPV{a}{}} = 0
\end{equation}
\noindent See~\cite{jereminov2016improving} for complete derivation of (\ref{eq:mpp}).

For $\PPDM{A}$, we form the following optimization problem to find the MPP, as there is no closed-form control equation for a non-uniform array:
\begin{subequations} \label{eq:mpp_algo}
\begin{equation}
    \begin{split}
        \max_{\VPV{p}{},\VD{p}{},\IOUT{p}{}} \Iarray{}\Varray{}\\
    \end{split}
\end{equation}
subject to:
\begin{equation}
    \begin{split}\label{eq:mc_cktPSE_hrtu}
         h(\VPV{p}{},\VD{p}{}, \IOUT{p}{}) = 0
    \end{split}
\end{equation}
\end{subequations}
\noindent The equality constraint $h(\cdot)$ in \eqref{eq:mc_cktPSE_hrtu} includes \eqref{eq:Iarr_sumIout}, \eqref{eq:Vgnd}, \eqref{eq:Varr}, \eqref{eq:perpanel_final_eq_1} and \eqref{eq:perpanel_final_eq_2} or \eqref{eq:node_2_w_diode} which represents the physics of the $\PPDM{A}$ of an  array with or without bypass diode.
\rev{Note that $\PPDM{A}$ requires per-panel parameters and operating conditions, and will collapse to $\SDM{A}$ under uniform conditions.}

%% file: Experiment_setup.tex
\noindent We compare the aggregated $\SDM{A}$ and $\PPDM{A}$ models using a real-world 10$\times$3 PV array composed of ET-M672BH395GL panels.
Panel parameters are derived from IV curve scans of on-site PV panels using standard methods \cite{duffie2020solar, Villalva2009}.
{We document the model parameters in the project's GitHub repository:~\url{https://github.com/Asang97/2026_solarPV_analysis}}.
\rev{The array we investigated is limited in size; thus the computational performance of $\PPDM{A}$ for utility-scale arrays remain to be studied.}

The array model explicitly accounts for bypass and blocking diodes between panels and strings, consistent with practical PV array configurations, to mitigate mismatch effects.
We model the bypass diode using the Shockley equation \eqref{eq:bypass_diode_current} with parameters tuned to a forward operating-point voltage of approximately 0.7 V at typical forward currents, while the blocking diode is assumed ideal.

\begin{figure}
    \centering
    \begin{subfigure}[t]{0.45\linewidth}
        \centering
        \includegraphics[width=\linewidth]{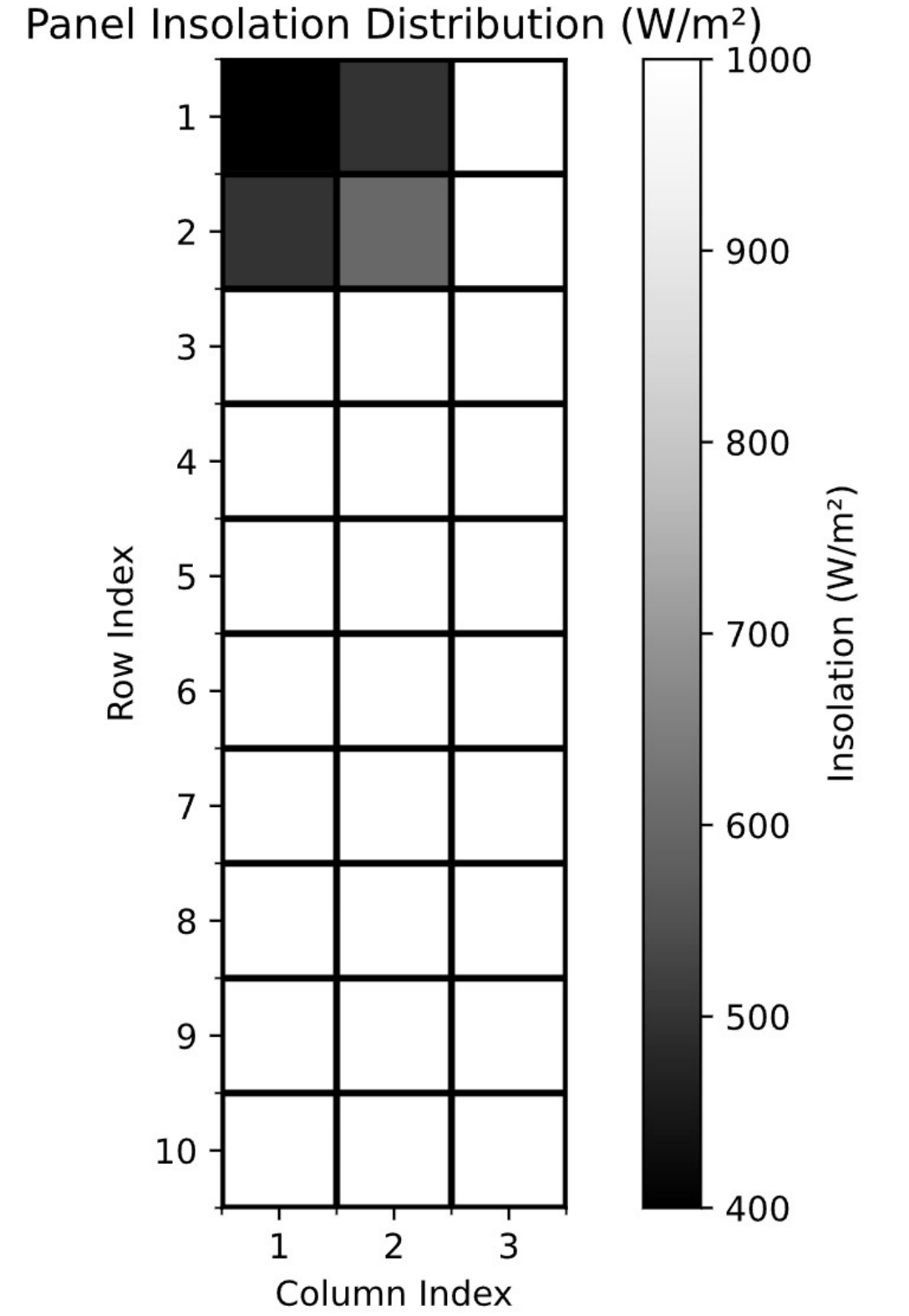}
        \caption{Partial shading scenario.}
    \end{subfigure}
    \hfill
    \begin{subfigure}[t]{0.45\linewidth}
        \centering
        \includegraphics[width=\linewidth]{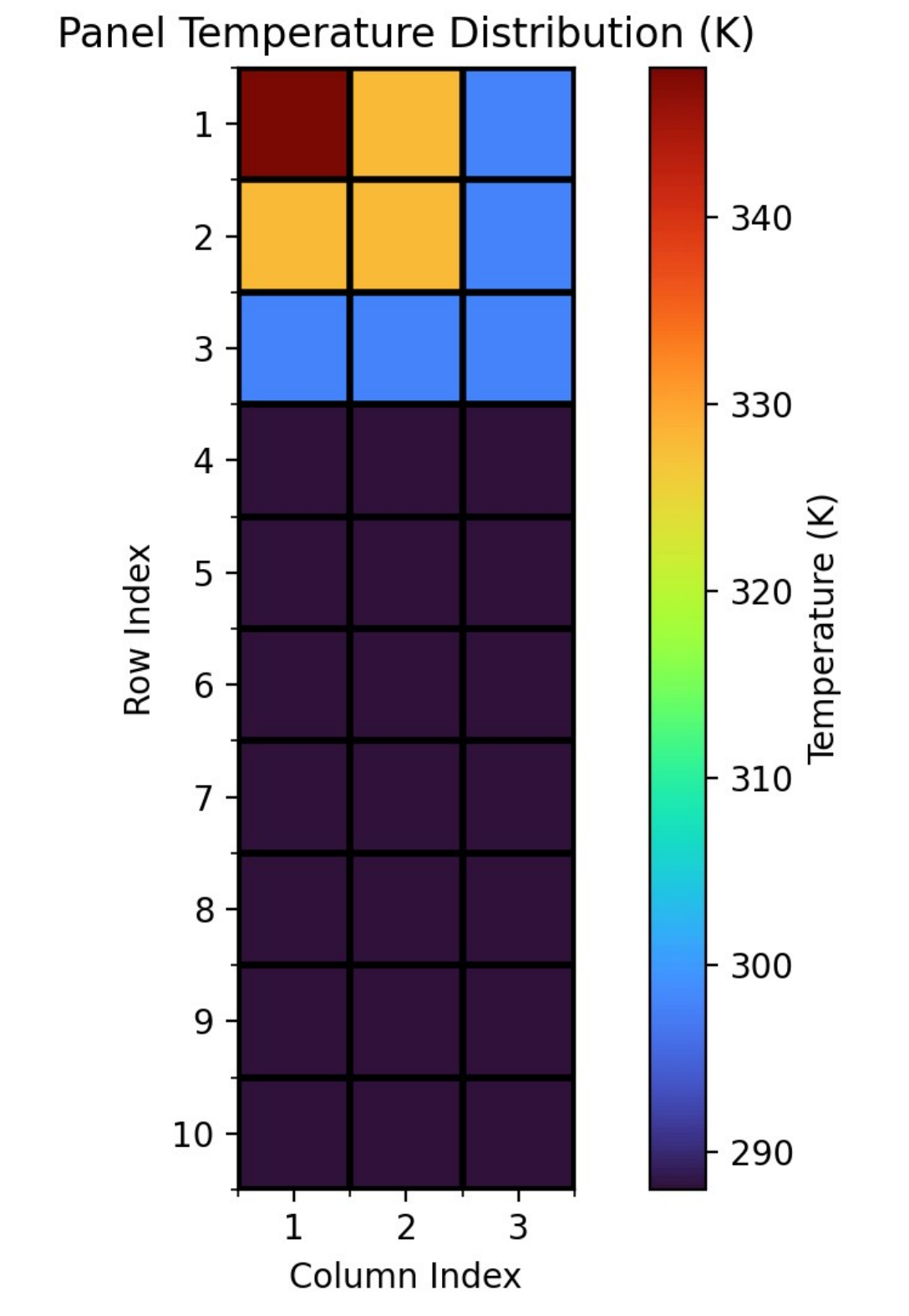}
        \caption{Hotspot scenario.}
    \end{subfigure}

    \caption{Non-uniform operating scenarios for a 10$\times$3 PV array:
    (a) partial shading at the top-left corner,
    and (b) hotspots in the upper rows.}
    \label{fig:non_uniform_scenarios}
\end{figure}

%% file: Experiment.tex
\noindent We evaluate the performance of $\SDM{A}$ and $\PPDM{A}$ under three operating scenarios:
(i) uniform panels (Scenario 1),
(ii) partially shaded panels (Scenario 2), and
(iii) panels with hotspots (Scenario 3).
For each scenario, we compare the two models in terms of (i) \textit{I-V} curve tracing and (ii) maximum power point tracking (MPPT) output.

\subsection{IV-Tracing Curve Results}
\begin{figure*}
    \centering
     \includegraphics[width=0.90\textwidth]{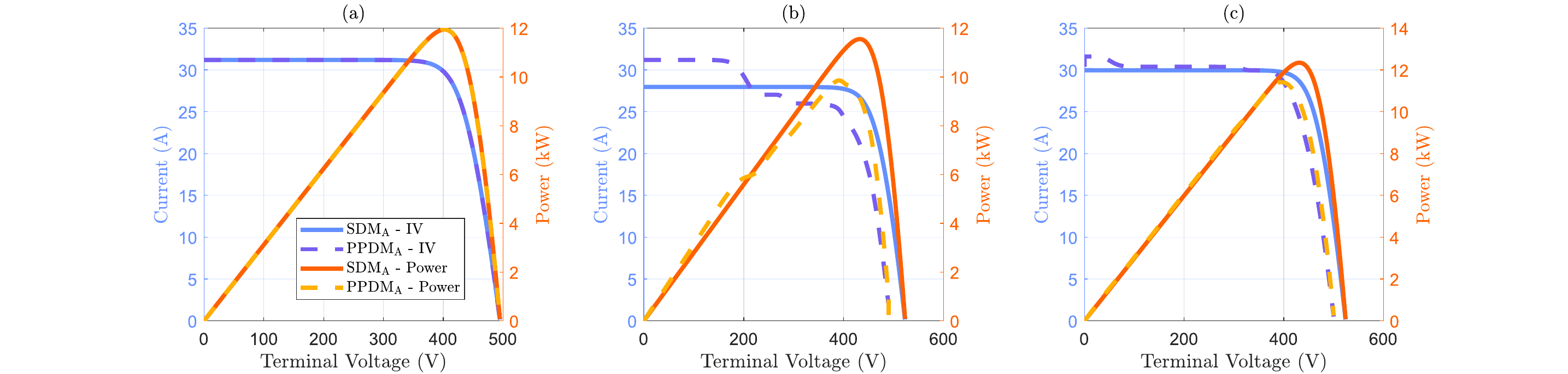}
    \caption{I-V and power curves of a PV array under:
    (a) uniform irradiance and temperature,
    (b) partial shading,
    and (c) hotspot conditions.}
    \label{fig:IV_Power_all}
\end{figure*}
\noindent \textbf{Uniform panels}:
Fig.~\ref{fig:IV_Power_all}(a) shows that when all panels in the array are uniform, the \textit{I-V} and \textit{P-V} characteristics of $\SDM{A}$ and $\PPDM{A}$ overlap.
This shows numerically and empirically that the aggregated model is lossless when uniformity holds.

\noindent \textbf{Partial shading condition}:
In this scenario, 26 out of the 30 panels operate under uniform irradiance of 1000~W/m$^{2}$; the remaining four panels in the top-left corner have irradiance levels of 400, 500, and 600~W/m$^{2}$ as shown in Fig.~\ref{fig:non_uniform_scenarios}(a).
We assume panel temperature is uniform across the array at 298~K.
We use the average irradiance of the array (933.3~W/m$^{2}$) as effective irradiance for $\SDM{A}$.

In this case, the \textit{I-V} and \textit{P-V} characteristics of $\SDM{A}$ and $\PPDM{}$ no longer overlap.
Fig.~\ref{fig:IV_Power_all}(b) presents the corresponding \textit{I-V} and~\textit{P-V} curves for both models.

\noindent\textbf{Panels with hotspots}:
In this scenario, the majority of the panels have a temperature ($T$) of 288~$K$, while nine panels located in the upper rows experience elevated temperatures of 348, 328, and 298~$K$ as shown in Fig. \ref{fig:non_uniform_scenarios}(b).
We assume the solar irradiance is uniform at 1000~W/m$^{2}$ for all panels.
For the $\SDM{A}$, we use average operating conditions of 1000~W/m$^{2}$ and 295.7~K for effective irradiance and cell temperature.
We documented the detailed procedure to compute the effective average irradiance and temperature values for $\SDM{A}$ in the project’s GitHub repository.
Under hotspot conditions, the \textit{I-V} and \textit{P-V} characteristics of $\SDM{A}$ and $\PPDM{}$ no longer overlap.
Fig.~\ref{fig:IV_Power_all}(c) shows the corresponding \textit{I-V} and \textit{P-V} curves for both models.

\subsection{MPPT Operating Point}
\noindent Table \ref{tab:MPP_summary_combined} summarizes the MPPT operating point for Scenarios 2 and 3 (in Fig. \ref{fig:non_uniform_scenarios}(a) and Fig. \ref{fig:non_uniform_scenarios}(b)).
We assume a higher-fidelity $\PPDM{}$ model is more realistic \rev{compared to $\SDM{A}$ because it preserves panel-level variations}.
\rev{We observe $\SDM{A}$} overestimates the power output by overestimating both the terminal voltage and the current of the array, with voltage being the larger contributor in percentage terms.
For numerical comparison, we compute the percentage \rev{discrepancy} between the power output of the two models during MPP by \eqref{eq:error}:
\begin{equation}\label{eq:error}
   \text{\rev{Discrepancy}}(\%) = 100\frac{P_\text{MPP}^\text{SDM} - P_\text{MPP}^\text{PPDM}}{P_\text{MPP}^\text{PPDM}}
\end{equation}
Similarly, we calculate the percentage \rev{discrepancy} of $V_\text{MPP}$ and $I_\text{MPP}$ between the two models.

\rev{Relative to $\PPDM{}$, $\SDM{A}$ exhibits discrepancies of 17.2\%, 10.7\%, and 6.1\% in power, terminal voltage, and current, respectively, under partial shading, and 5.5\%, 3.4\%, and 1.7\% under hotspot conditions.
These results show that $\SDM{A}$ introduces scenario-dependent discrepancies in the MPPT operating point and predicts higher power under non-uniform conditions due to its uniformity assumption.}

\begin{table}[t]
\caption{MPP summary of $\SDM{A}$ and $\PPDM{}$ under different non-uniform conditions}
\label{tab:MPP_summary_combined}
\centering
\setlength{\tabcolsep}{4pt}
\resizebox{\columnwidth}{!}{
\begin{tabular}{c|ccc|ccc}
\hline
 & \multicolumn{3}{c|}{PSC} & \multicolumn{3}{c}{Hotspot} \\
Model
 & $P_\text{MPP}$ (W) & $V_\text{MPP}$ (V) & $I_\text{MPP}$ (A)
 & $P_\text{MPP}$ (W) & $V_\text{MPP}$ (V) & $I_\text{MPP}$ (A) \\
\hline\hline
$\SDM{A}$   & 11552.6 & 432.2 & 26.7 & 12039.3 & 408.8 & 29.4 \\
$\PPDM{}$   &  9858.8 & 390.6 & 25.2 & 11416.6 & 395.3 & 28.9 \\
\hline
\end{tabular}}
\end{table}

%% file: Conclusion.tex
\noindent
We draw the following conclusions based on the presented analysis:
(i) $\PPDM{A}$ captures complex nonlinearities that $\SDM{A}$ cannot represent when tracing \textit{I-V} characteristics at the array terminal for different operating conditions.
\rev{(ii) $\SDM{A}$ predicts MPP power values 17.2\% and 5.5\% higher than $\PPDM{A}$ under partial shading and hotspot conditions, due to its uniformity assumption.}